\documentclass[11pt]{article}

\usepackage{times}
\usepackage{graphicx}

\normalfont

\newcommand{\capb}{\mathbf{B}}
\newcommand{\drm}{\,\rm d}

\begin{document}
%\begin{frontmatter}

% The preamble begins here.
%\pretitle{DRAFT}

\title{Macrodynamics of users' behavior in Information Retrieval}

\author{Daniel Sonntag\thanks{DFKI, German Research Center for
    Artificial Intelligence, 66123 Saarbruecken, Germany}, and Rom\`an
  R. Zapatrin\thanks{Informatics Dept, The State Russian Museum,
    In\.zenernaya 4, 191186, St.Petersburg, Russia}}

\maketitle

\begin{abstract}
We present a method to geometrize massive data sets from search
engines query logs. For this purpose, a macrodynamic-like
quantitative model of the Information Retrieval (IR) process is
developed, whose paradigm is inspired by basic constructions of
Einstein's general relativity theory in which all IR objects are
uniformly placed in a common Room. The Room has a structure similar
to Einsteinian spacetime, namely that of a smooth manifold.
Documents and queries are treated as matter objects and sources of
material fields. Relevance, the central notion of IR, becomes a
dynamical issue controlled by both gravitation (or, more precisely,
as the motion in a curved spacetime) and forces originating from the
interactions of matter fields. The spatio-temporal description
ascribes dynamics to any document or query, thus providing a uniform
description for documents of both initially static and dynamical
nature. Within the IR context, the techniques presented are based on
two ideas. The first is the placement of all objects participating
in IR into a common continuous space. The second idea is the
`objectivization' of the IR process; instead of expressing users'
wishes, we consider the overall IR as an objective physical process,
representing the IR process in terms of motion in a given
external-fields configuration. Various semantic environments are
treated as various IR universes.
\end{abstract}

% \begin{keyword}
% IR Models \sep  spatio-temporal description  \sep IR universes
% \end{keyword}
% \end{frontmatter}
%\thispagestyle{empty}
%\pagestyle{empty}

\section*{Introduction}

The goal of this paper is to provide a framework in which to compare
and introduce new Information Retrieval methods, rather than to propose a
particular retrieval strategy. In order to enhance the capabilities of
search engines, we need to know how well the engines satisfy the user
requests. We try to answer this question by trying to understand the
user or user group behavior.

New insights can be gained by mining search patterns, or as
complementary approach, by visualizing the click streams in an
intelligent way so that an expert can make sense of the structures he
detects in the visualizations (visual data mining). Especially in the
case of large data sets, a method of geometrization from search
engines query logs is very much in demand. To Manage huge data
corpora, a proper theory for its description is required. Once the
data are represented in the database, two different types of queries
can be started, resulting in very different query processing
stages. The interpretability of the returned results is different as
well. In Data Retrieval, only exact matches to a query are considered,
whereas in Information Retrieval, documents with a certain probability
of relevance to the query are searched.  Information Retrieval queries
are, technically speaking, $k$-nearest-neighbor queries with
similarities adopted for the specific information need.

In order to build \emph{an} Information Retrieval theory, structured
data are to be represented in the data retrieval context in some way.
The data are treated as discrete by nature, but this does not
imply that they have to be put in a discrete environment (consider, for example,
an appropriate analogy in the theory of solid states or
liquids: everyone knows that they are composed of discrete
molecules). However, continuous geometrical methods proved high
efficiency and predictive power. This, in turn, is the result of a
crucial simplification of the model by disregarding its
micro-details.

The vector model was the first considerable step in this direction. It
introduced vector spaces (which are spanned on terms or their
generalizations) and treated documents and queries uniformly as
vectors in the same space. These spaces are still discrete.

Theoretical physics and, more generally, the physical world remain
a source of inspiration for computer scientists \cite{GBM}. The
first really continuous model was suggested by C. J. Keith van
Rijsbergen who introduced Hilbert spaces for this purpose (as in
quantum mechanics). By nature, quantum mechanics is a genuine
combination of continuous and discrete. In van Rijsbergen's model,
the relevance becomes the angle or the distance between appropriate
continuous vectors.  This is an effective illustration of the idea
of a \emph{quantum-like} description \cite{khrennikov-2007}: it has
nothing to do with its roots in quantum physics, nonetheless, it
efficiently uses its mathematical language and results.

We, however, should strike a new path. Two major
fundamental theories in modern physics exist which are mutually exclusive to
a great extent: Quantum Mechanics and General Relativity. The former
mostly deals with the microworld, whereas the latter deals with cosmic
distances. Our everyday intuition is in between  and called
classical physics; both theories admit the so-called classical
limit. Vector models are based on a quantum-mechanical, linear vector
space paradigm which plays the r\^ole of a Room to store data and pose
queries.

\medskip

Our basic idea is the following. We consider a smooth continuous
manifold $\capb$, and call it Information Retrieval space.  Note that
$\capb$ is neither a document space, nor a query space; instead, it
has a more fundamental and unstructured nature. It may be thought of
as the set of all transmitted bits. The elements of $\capb$ are all
the same; they have no structure. This is a complete analogy to the
points of spacetime, or the configuration space in physical theories.

For the time being, let us return to van Rijsbergen's geometrical
model.  The documents in his approach are vectors, but, if we look at
the model in more detail, we see that if we multiply a vector by a
number, we do not get a new document. As in quantum mechanics, only
unit vectors are of physical (operationalistic) meaning.  These
vectors, in turn, form a smooth manifold --- the unit sphere in the
appropriate vector space. Therefore, even in van Rijsbergen's
approach, curved spaces are already implicitly used as document
spaces.  In quantum mechanics, quantum dynamics have been successfully
described as classical Lagrangian mechanics on unit vectors
\cite{ashtekar-schilling}.  It should be mentioned that all this has
no direct relevance to our approach, but shows that what we suggest is
a natural development of standard, conventional approaches.

To be more specific, we treat $\capb$ as an analogon to physical
spacetime. We place both documents and queries into $\capb$, providing
them with both temporal and spatial dimensions. As a consequence, the
idea that a document may change in time is automatically incorporated
in the theory. The second consequence is that the static documents,
and those generated on-the-fly, are described as entities of exactly
the same nature, differing only in `shape' in our IR spacetime
$\capb$.

\section{Information retrieval as dynamics}\label{sinforet}

In this section we develop one of the idea highlighted in the
beginning, namely the objectification of the IR process. Information
Retrieval is commonly treated as an analogy to data search: there is a
user with a (more or less) definite goal wishing to gain this or that
knowledge from the retrieved information.

We suggest an alternative approach: When we are speaking about a huge
community of users, we no longer treat their behavior as
intelligent. This contrasts with the viewpoint of `intelligent crowd
behavior', but the community of users in our setting is a large
collection of autonomous individuals rather than a crowd, and we dwell
on their average behavior. This gives us the right to shift the focus
of our research from the task of finding a good way to satisfy users'
requests to the task of analyzing typical user behavior. From this
perspective, a typical user of a search environment is not more
intelligent than an elementary particle or a molecule, and we may
apply the good old principle of least action, which stems from the
work of Fermat and Euler. They proved its efficiency by providing
simple and strongly predictive models. The power of the least action
principle is that we do not have to make any difference between users
and resources---we are free to include anything we like in a uniform
way to describe the dynamics.

\paragraph{IR environment.} This notion is informal;  we need it to link the
mathematical model with practical situations. Within a mathematical
model, the IR environment is specified by the IR space $\capb$, a
collection of effective fields on it, and the Lagrangian (which is a
concrete expression for the action). As soon as all this is
specified, the IR process itself is represented by trajectories in
$\capb$ which show the behavior of users.

The standard IR paradigm treats the IR process as a search. That is,
the initial condition is posing a query, then, according to this or
that formula, the indexed documents are ranked. Subsequently, the
results are delivered to the user according to the ranking. But,
typically, the user never makes a single query and the process is
usually progressive. After parsing the results and considering their
relevance, the user poses further queries, repeating the process
iteratively.

Our suggestion is to get above these particularities. We replace the
notion of relevance feedback by that of least action\footnote{A
  similar approach appears in the ostensive model of information
  retrieval \cite{campbell}. Within this model, there is an implicit
  unobservable entity---state of knowledge, or awareness of a
  user---and the behavior of the user is interpreted as the change of the
 user's knowledge. The user acquires knowledge after
  performing a certain action.}. This can be drawn as follows
\begin{center}
\includegraphics[width=0.5\textwidth]{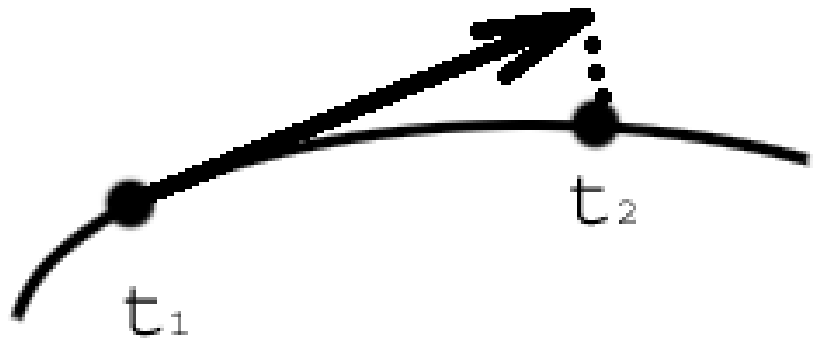}

Figure 1. A point on IR surface together with users' intention
vector.
\end{center}
and interpreted as geodesic motion. The dynamics replace the notion
of relevance, and the displacement of a user from point $t_1$ to
point $t_2$ is what replaces relevance feedback making it, in a
sense, a \emph{relevance feedforward}.

\paragraph{Users.} We should not treat users literally as persons.
 In our setting, a user is just an entity which pursues a particular
 goal.  This means that a single physical person can represent a
 number of simultaneously acting users and, conversely, there may be a
 group of people whose retrieval behavior looks like that of a single
 user from the outside.  Later we shall examine the problem of
 detecting users and user clickstreams in more detail.

\section{An outline of differential geometry.}\label{sdiffgeo}

The aim of this section is to present the basic geometric ingredients
for our model and introducing the notation. We start by presenting the
basic notions of a smooth manifold, followed by the Riemannian and
pseudo-Riemannian metrics and geodesics.

A \emph{smooth manifold} $A$ is an analogon to a curved surface with
the difference that it is considered \emph{per se}, not merged into
any outer space. \emph{Metric} in differential geometry has a double
meaning, a global and a local one. The global metric is a distance
function ascribed to any pair of points and satisfying the triangle
inequality. Locally, a metric is a nondegenerate quadratic form
$g(u,v)$ which defines the scalar product for any pair of tangent
vectors. If the quadratic form $g$ is positive-definite, the
appropriate metric is called \emph{Riemannian}. However, the metric is
called \emph{pseudo-Riemannian} when squared lengths of vectors maybe
both positive and negative (and zero as well). The latter is the
mathematical ground for special relativity as spacetime is assumed to
define such a metric; that is, time is treated as complex-valued
distance.

A \emph{geodesic line} is an analogon to a straight line on a plane.
This is the locally shortest curve, shortest with respect to the
defined metric on the manifold $A$. For instance, circumstances are
geodesics on a sphere. The explicit formula for geodesics is as
follows: Given a metric $g$, fix a coordinate system, then $g$ takes
matrix form $g=g_{ik}$, each matrix entry is a  function defined on
the manifold $A$. Combine their derivatives, introducing the
coefficients:
  \begin{equation}\label{egamma}
    \Gamma^j_{kl}
    \;=\;
    \frac{\partial g_{lk}}{\partial x_j}
+    \frac{\partial g_{jk}}{\partial x_l} -
\frac{\partial g_{kl}}{\partial x_j}.
  \end{equation}

\noindent When the coefficients are calculated, the equation of
geodesic motion $x(t)$ along the manifold $A$ is the following
second-order differential equation:
\begin{equation}\label{egeodesic}
\ddot{x}^j
\;=\;
    \Gamma^j_{kl} \dot{x}^k \dot{x}^l.
\end{equation}
Here, the dot above indicates the derivative over a parameter counting
the points of the trajectory $\dot{x}^j=\partial x^j/\partial t$,
and the summation over repeated indices is assumed. This means the
expression $\Gamma^j_{kl}\dot{x}^k\dot{x}^l$ is in fact
$\sum_{k,l}\Gamma^j_{kl} \dot{x}^k \dot{x}^l$.

\medskip

In a local sense the metric is connected to the global one as
follows. Given a curve $x(t)$, its length is given by the integral
\begin{equation}\label{elength}
    \int_{0}^{T} g(\dot{x}(t),\dot{x}(t)) \drm t
    \;=\;
    \int_{0}^{T} g_{jk}\dot{x}^j(t),\dot{x}^k(t) \drm t
\end{equation}

\noindent where (i) the summation is carried over repeated indices
and (ii) $g_{ik}=g_{ik}(a)$ are functions, defined at each point
$a\in A$.

\paragraph{Dynamics.} As stated above, we replace the study of
users' needs with the study of users' behaviors in a way analogous to
the study of a deterministic physical processes. For that, we
introduce the notion of \emph{action} as a function which evaluates
every curve (the basic example of action is the length of the
curve). Given an action, we then use the well-known fundamental
physical principle of least action: Among all possible trajectories it
happens that (only) those yielding the minimum to the action are
realized. In our approach, all the variety of evaluating relevance is
assumed to be hidden in the calculation of the action.

\paragraph{Describing manifolds.} How can we
generally describe infinite, continuous objects?  This immediately
brings us to the question of how we can describe a function which, in
turn, has commonly accepted answers. We treat certain sets of
functions as elementary and construct new functions from them using
elementary operations. What is elementary? This is a matter of the
particular setup of the problem to be defined individually.

In our case we are going to deal with regular geometrical objects and
simply treat smooth manifolds as surfaces in Euclidean space, defined
by appropriate smooth functions. In particular, when we reconstruct
smooth surfaces from discrete data, we use standard approximations
such as the mean square method with respect to Euclidean distance.

\section{Building IR spaces}\label{sbuilding}

We begin by drawing an analogy between IR spaces and differential
geometry, in the context of smooth manifolds. When we just say `given
a manifold' this still means nothing unless we specify it. We have
already presented a method of building IR spaces by representing them
as graphs of smooth functions.  Another way to represent IR spaces is
to specify a manifold by describing the set of all smooth functions on
it. (These sets are different. For instance, any such function on a
circle attains its maximal value, which is no longer the case for a
straight line.) An algebra is a linear space with an extra operation
of multiplication. One can easily observe that, given a space, the set
of all functions on it is closed under pointwise addition and
multiplication. That is, the set of functions is a linear space
equipped with an extra operation of multiplication, such spaces are
called \emph{algebras}.

\subsection{Dimensionality reduction}\label{sdimred}

A dimension can be defined as one of a number of parameters needed to
describe an object. This may sound abstract, but there are parallels
with our everyday experience. A cake recipe, for example, may be
defined by the amount of the various ingredients in grams. If one
writes down the amounts of flour, sugar, butter, eggs, and baking
powder in the form (200, 100, 80, 20, and 3), then this representation
contains the most important information. So, there is essentially
nothing complicated with five dimensions from a common sense
point-of-view (one may even use this example to explain the vector
space model for IR).

Mathematical methods of dimensionality reduction can be used for
feature transformation. Feature selection, for example, focuses on
uncovering subsets of variables predictive of a prespecified target
variable. In our context, dimensionality reduction comes into
consideration when we want to control the number of parameters for the
results of visualization.

The dimension is one of the main properties of linear spaces; it may
finite or infinite. In the case of an algebra of functions on a
manifold, the dimension is infinite. What does that mean? Suppose we
would like to specify a straight line. We might consider the linear
space of polynomials, treated not as functions, but defined formally,
as, say, formal series. The dimension of this space is obviously
infinite as nobody limits the degree of polynomials. In the meantime
we know that the space, on which these polynomials are defined, is
just a straight line, a one-dimensional object! And it is completely a
matter of our choice which of the descriptions of the straight line we
prefer: either functional and infinite-dimensional, or geometrical and
one-dimensional.

\medskip

After that, we can present to our basic suggestion. By analogy with
algebras we see that we may define the IR space in terms---thus making
it huge-dimensional---or, rather, observe some `massive regularities'
and define the IR space geometrically, as an abstract manifold
$\capb$.  The terms will then become functions on $\capb$, exactly as
in differential geometry.

Dimensionality reduction is one of the key features of our approach.
This is reason why we do not treat terms as basic objects: the
appropriate vector space would have an immense dimension. What we
suggest is a kind of holographic approach. Its closest analogy in
image processing is the JPEG format. If we draw an analogy with image
processing, terms will be a counterpart to pixels, vector models are
then similar to the BMP format; we parameterize the search space by
holistic patterns.

\subsection{IR Space from discrete skeleton}\label{suserstraj}

In this section we will dwell on the first basic principle of our
techniques: merging everything---users, queries and data---into a
single space.

Return to equation \ref{egeodesic}. It is of second order, that
is, in order to specify its particular solution, we must specify the
initial conditions which are the initial position $x(0)$ and the
initial `intention' $\dot{x}(0)$. A typical user clickstream will be
represented as a line, whereby the points of the line $x(t)$ are
associated with the state of knowledge the user has gained from
interpreting the retrieved information until that point.

Next, let us specify \emph{what} are we going to visualize. The object
of our inquiry is the \emph{IR semantic environment}, which consists
of a typical community of users with specific needs, using certain
information or knowledge retrieval techniques. In fact, this
requirement is not crucial, we may take a random collection of users,
and even carry out its visualization, but the point is that this
visualization will remain a thing in itself. If, conversely, we
determine some common features of the team of users, we may afterwards
vary the parameters of the problem and the obtained visualization may
give us an immediate tracking of the results. However, at present, we
may not put \emph{a priori} restrictions on the contents of the
environment.

We study the behavior of an IR environment by analyzing the logs of
user querying activities. Let us first produce the `flesh' of IR
space. Its elementary constituent, a point, is a click: a query/HTTP
request together with a body response (HTML page accessed by a result
link -- this way we do not take into account broken links).

\paragraph{Step 1. Extract the clickstreams.} A clickstream is a progressive,
 `continuous' sequence of user queries and responses which have a
definite start and end. The end of a clickstream is marked by a
breach in the continuity of the requests. What does `continuous'
mean? To specify it we need a distance function between points. This
distance is beyond our exploration in this paper, but we may use one
from, say, vector space model.  The result of Step 1 is a collection
of clickstreams, an ordered sequences of points:
\begin{center}
\includegraphics[width=0.4\textwidth]{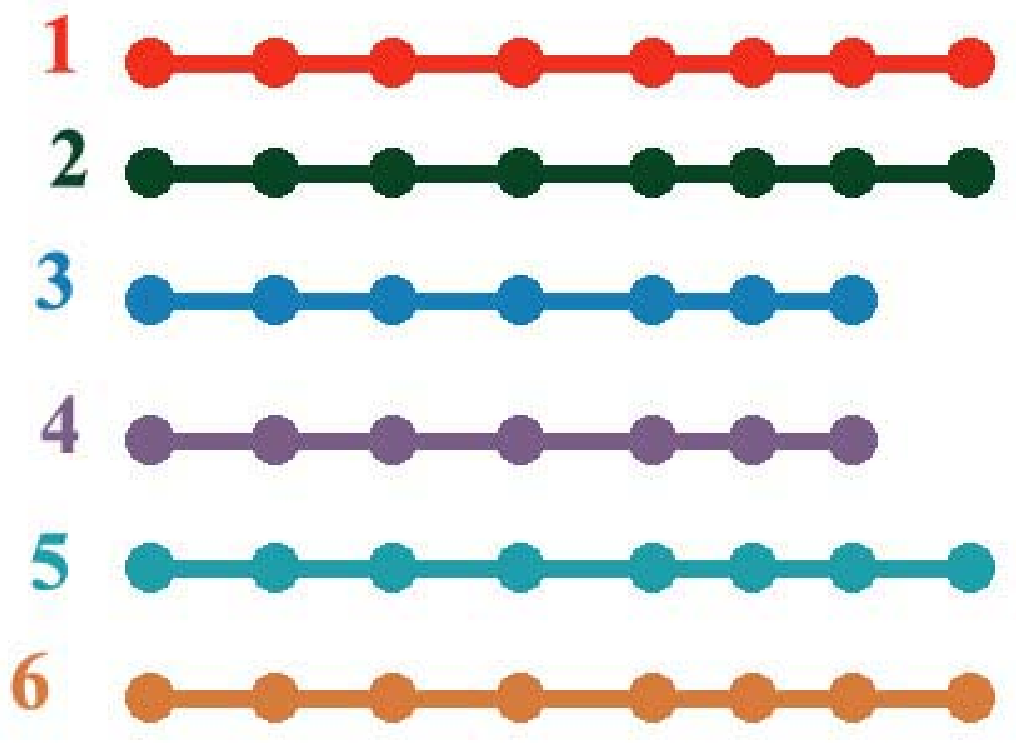}

Figure 2. Points on clickstreams are ordered by timestamps of
clicks.
\end{center}

\paragraph{Step 2. Creating the discrete pre-space $\capb$.} At this step, our input is
a collection of clickstreams. Their points are ordered and we know
the distances between them. We assume that we use certain distance
between points of the threads, and therefore, any particular
relevance formula can be applied. There is a well-defined distance
between the neighbor points of each thread. That means, beside
order, clickstreams acquire metric:
\begin{center}
\includegraphics[width=0.5\textwidth]{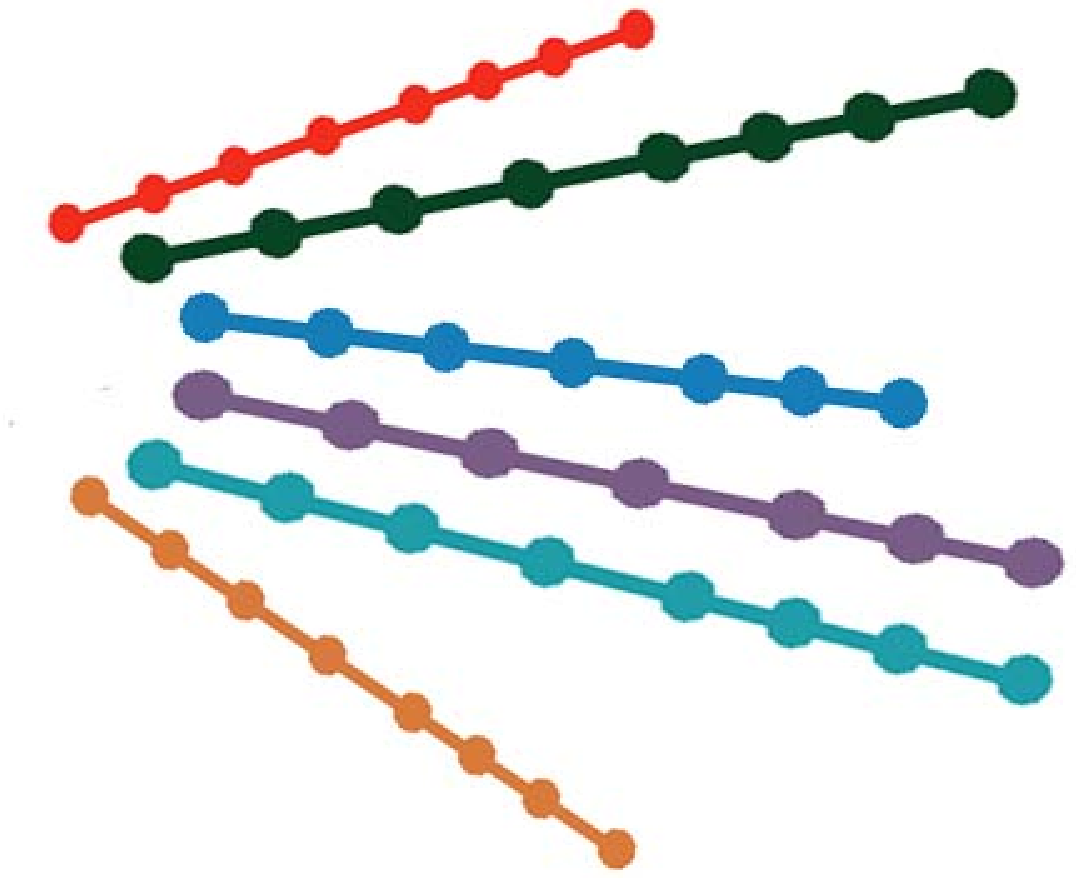}

Figure 3. Clickstreams acquire metrics.
\end{center}

\noindent Now we make a layered structure. We start with points with
label 0 (this will be a starting layer), and, using the same
distance function,  place them as points on a metric space. Then we
pass to label 1, and form the same discrete metric space, and so on.
As a result, we have a sequence of layers labeled $0,1,\ldots$,
forming altogether a discrete metric space:
\begin{center}
\includegraphics[width=0.5\textwidth]{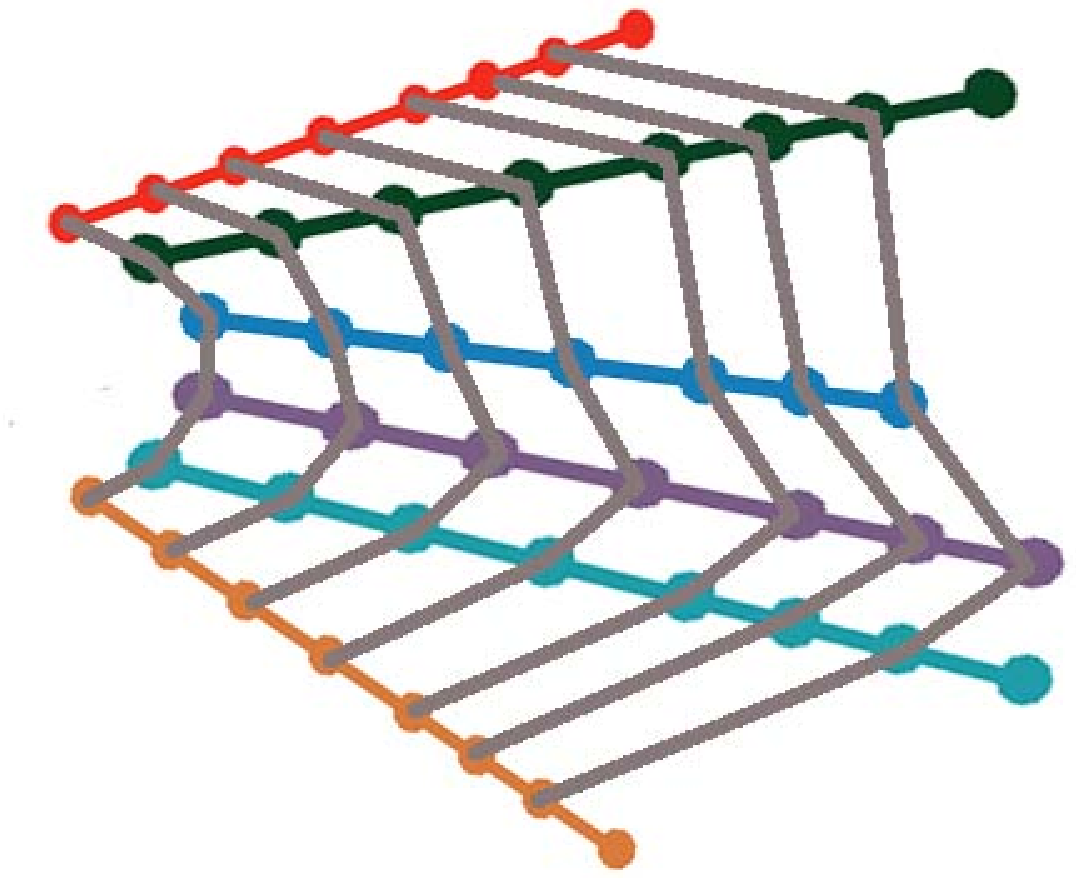}

Figure 4. Creating transversal layers.
\end{center}

\paragraph{Step 3. Geometrization and dimensionality reduction.} At this
step we continue binding the threads and complete the skeleton with
the `bones' linking nearest neighbors, now irrespective of the
thread, to which they belong
\begin{center}
\includegraphics[width=0.5\textwidth]{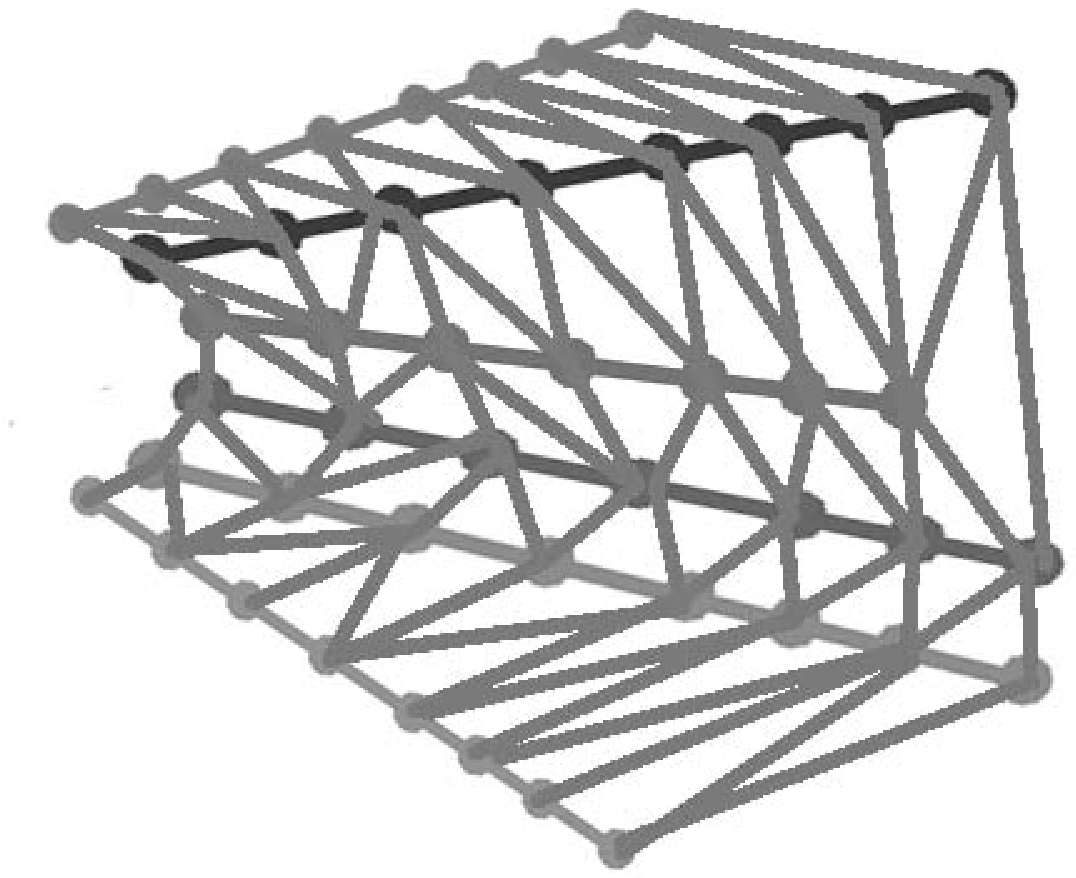}

Figure 5. Forming the discrete skeleton.
\end{center}

\noindent we choose, or set up by force, the dimension $n+1$ of our
IR space. (Since it has a spatiotemporal structure, we reserve one
dimension for the temporal parameter and $n$ for `spatial'.)  Once
$n$ is chosen, we project each layer on an $n$-dimensional space.
Technically, this can be done as follows: when the dimension $n$ is
fixed, we form cells of $n+1$ neighboring points for each point and
then treat each such collection of points as a simplex (simplex is a
generalization of a triangle, pyramid and so on). So, we form a
foliation, labeled  $1,2,\ldots$, together with threads, which we
retained from Step 1. Finally, we treat the resulting space as IR space.
\begin{center}
\includegraphics[width=0.7\textwidth]{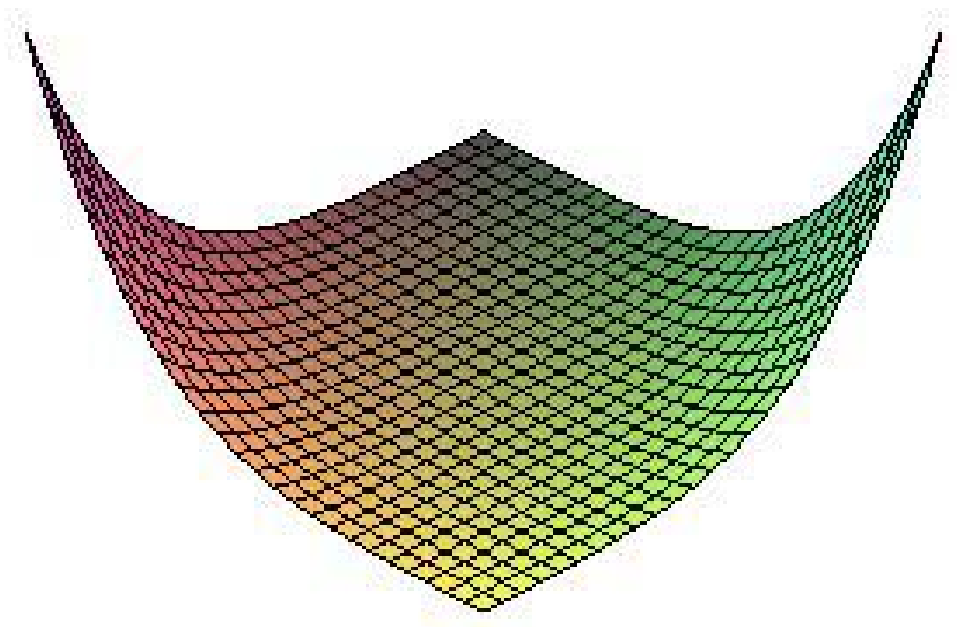}

Figure 6. IR space is built.
\end{center}

\section{Possible applications}\label{sappl}

Here we give a brief overview of the potential benefits of the
techniques we introduce.

\paragraph{IR spaces as comparison tools.}
Now, how can our geometrical picture serve as a comparison tool
between different IR environments? Suppose we have a kind of
contest. There are, say, two search environments and two similar teams
of users with the same tasks and wishes. After some time we may
represent the results of the contest for each environment as a
geometrical picture, i.e., a manifold and a collection of users'
trajectories on it. Since the teams of users are similar, we may put
them into a correspondence and thus establish a mapping between the
two manifolds.

This comparison can also be viewed from a physical perspective.
Suppose we have carried out such an experiment with a team of
observers.  Then, we change the circumstances and a new set of
relevant documents emerges and is indexed. As a consequence, the
behavior of users will also have changed.  This result has a direct
physical analogy: Suppose we have a cloud of test particles and we
record their trajectories. Then, a massive body emerges in the
neighborhood. As a result, the trajectories will be biased.

In general, this representation is not a way to judge which IR
environment is better; rather, it is a way to put them together and
visually compare them, thus making it an instance of visual data
mining in the sense of visual pattern recognition or the like.

\paragraph{Geometrodynamics and relevance feedback.} Our visualization
method can also be used as a relevance feedback
tool.  We may use it to modify the relevance-distance function. The
idea behind is the following: Suppose we look at a typical picture
of users' behavior and discover that there are sharp peaks on the IR
surface. What does that mean? Users typically make big jumps:
\begin{center}
\includegraphics[width=125mm]{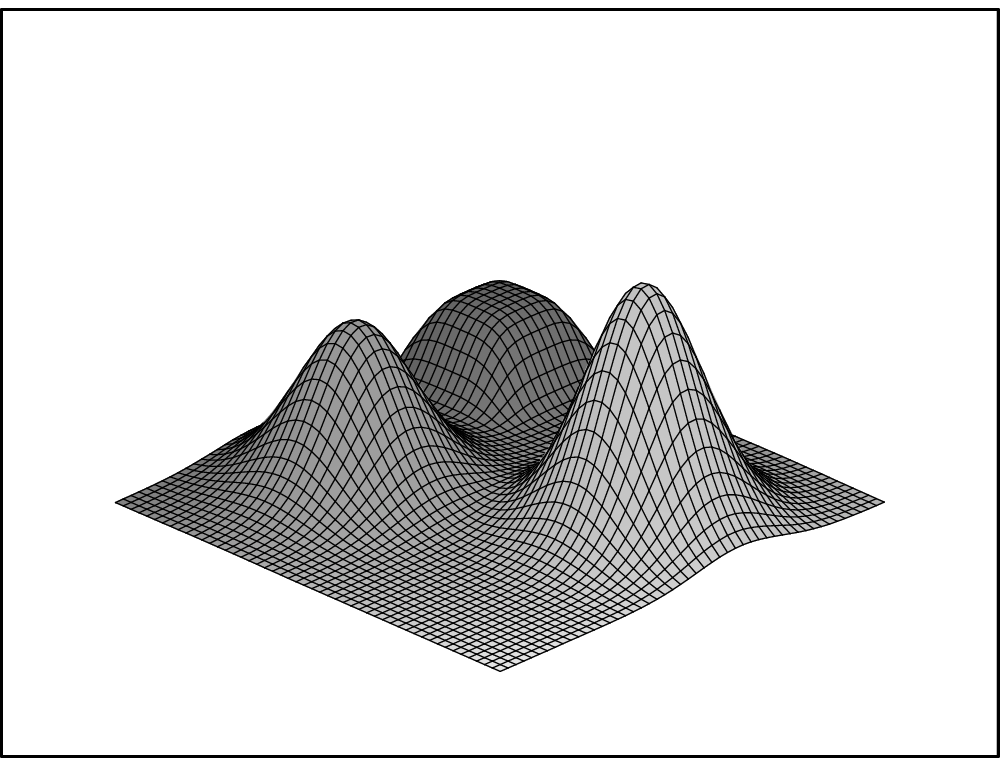}

Figure 7. High peaks on IR surface.
\end{center}
From this we may infer that our distance is not adequately
calculated and this will force us to correct the ranking formula.

Here, our goal is to make the surface more smooth and less lumpy,
according to the requirement to make the ranking function more
consistent with the users requests and their evaluations of the
results the retrieval. It works as depicted in Figures 7, 8: the
smoother is the surface, the better the IR is organized and the
simpler it is for users to achieve their goals.
\begin{center}
\includegraphics[totalheight=25mm,width=125mm]{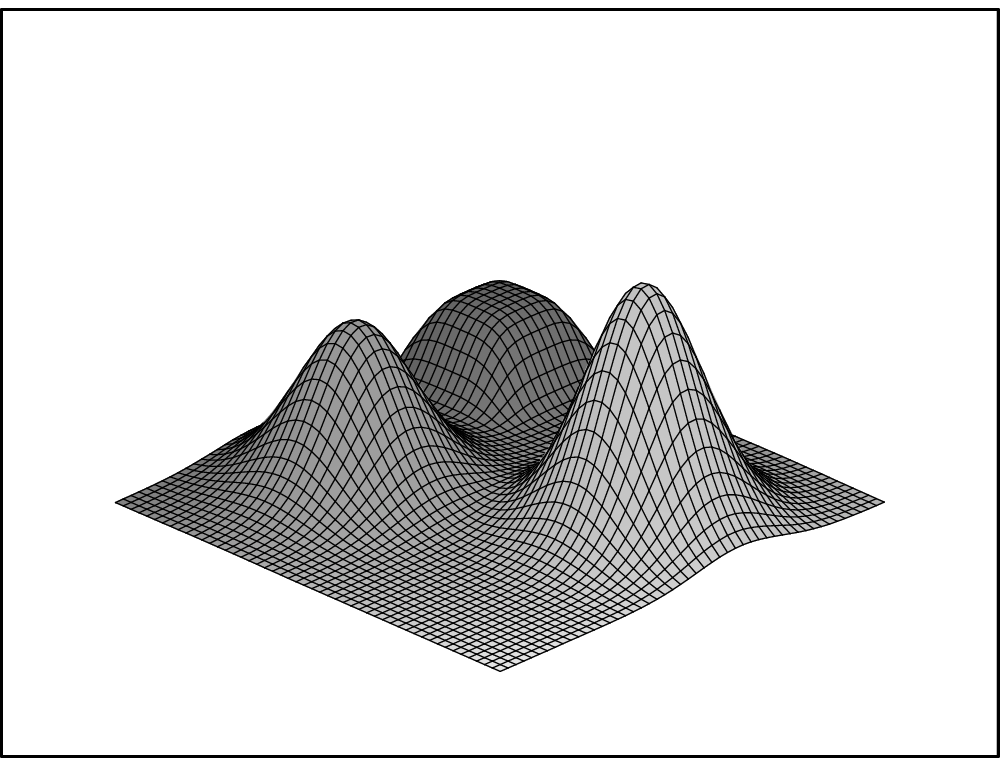}

Figure 8. Lower peaks after redefining the ranking function.
\end{center}

\section{Related works.}

\emph{If I have seen further it is only by standing on the shoulders
  of Giants.- I.Newton.} In this section we summarize and compare some
previous works and point out their relevance to our model.

Keith van Rijsbergen\cite{geometrikeith} has already been mentioned in
the introduction. Iain Campell\cite{campbell} developed an ostensive
model of IR. Duality issues were studied by R.Rousseau and Leo Egghe
\cite{rousseau97}.

In this paper the duality relation between documents and queries, and
between indexing and retrieval are studied.  This is an important step
towards merging all of these objects in a unique space. Recently, a
gravitation-based model (GBM) of IR was proposed, where
\emph{relevance} is treated as Newtonian gravitation between a query
and a document.  This provides not only a holistic view, but also a
mathematical background to deduce particular ranking formulas. In
particular, the famous Okapi BM25 formula
\begin{equation}\label{ebm25}
\mbox{\rm score}(D,Q) \;=\; \sum_{i=1}^{n} \mbox{\rm IDF}(q_i)\cdot
\frac{f(q_i,D)\cdot (k_1+1)}{f(q_i,D)\,+\,k_1\cdot\left(
1-b+b\cdot\frac{|D|}{\mbox{avgdl}}\right)}
\end{equation}
is naturally derived within their approach. GBM represents documents
as cylindrical objects and considers only attraction between documents
and point-like queries, according to the Newtonian gravity law. These are, however,
advantages over GBM and argumentations in favor of our model:

\begin{itemize}
\item Right from the beginning one sees that the Newtonian formula for
  gravity is too rigid and cannot properly capture the subtleties of
  relevance.
\item In order to adjust the function $f$ in \ref{ebm25} properly,
  the authors of GBM suggest replacing the Newtonian quadratic law
  with a different one, varying the power of the distance. This
  immediately destroys the beauty, simplicity, and self-consistence of
  the Newtonian world.
\item Our suggestion is different. Instead of modifying the law of
  gravity, we modify the geometry of the space, but leave the laws
  intact---exactly as it is done in Einstein's General Relativity.
\end{itemize}

An introduction to the possible applications, the topics of
interpreting public search queries can be found in \cite{spinketal01};
\cite{jansen01} provide a review of web searching studies, and
\cite{kammenhuberetal06} address the difficulties when processing web
search clickstreams.

\section*{Conclusion}

We presented a method to geometrize massive data sets from search
engines query logs. For this purpose, a cosmological-like quantitative
model of the Information Retrieval (IR) process has been developed,
where documents and queries are treated as matter objects and sources
of material fields.

One of the peculiarities of our approach is that we practically do
not use and do not consider terms as basic entities. We do that
deliberately in order to simplify the construction in some sense.

\paragraph*{Acknowledgments.}
Daniel Sonntag is supported by the THESEUS Programme funded by the
German Federal Ministry of Economics and Technology
(01MQ07016). Rom\`an Zapatrin acknowledges support from RBRF, grant
(07-06-00119-a).

\end{document}